\documentclass[prl,superscriptaddress,floatfix,twocolumn]{revtex4}
\usepackage{amsfonts}
\usepackage{stmaryrd}
\usepackage{bbm}
\usepackage{mathrsfs}
\usepackage{tipa}
\usepackage{amssymb}
\usepackage{txfonts}
\usepackage{graphicx}
\usepackage{dcolumn}
\usepackage{epstopdf}
\usepackage{array}
\usepackage{color}
\usepackage[colorlinks, linkcolor=blue, anchorcolor=blue, citecolor=blue]{hyperref}

\newcommand{\PreserveBackslash}[1]{\let\temp=\\#1\let\\=\temp}
\newcolumntype{C}[1]{>{\PreserveBackslash\centering}p{#1}}
\newcolumntype{R}[1]{>{\PreserveBackslash\raggedleft}p{#1}}
\newcolumntype{L}[1]{>{\PreserveBackslash\raggedright}p{#1}}
\renewcommand{\textcolor}[2]{#2}

\begin{document}

\newcommand*{\cm}{cm$^{-1}$\,}

\title{\textcolor{black}{Strongly Entangled Kondo and Kagome Lattices and \textcolor{red}{the Emergent Magnetic Ground State} in Heavy-Fermion Kagome Metal YbV$_6$Sn$_6$}}

\author{Rui Lou}
\thanks{lourui09@gmail.com}
\affiliation{Leibniz Institute for Solid State and Materials Research, IFW Dresden, 01069 Dresden, Germany}
\affiliation{Helmholtz-Zentrum Berlin f{\"u}r Materialien und Energie, Albert-Einstein-Stra{\ss}e 15, 12489 Berlin, Germany}
\affiliation{Joint Laboratory ``Functional Quantum Materials" at BESSY II, 12489 Berlin, Germany}

\author{Max Mende}
\affiliation{Institut f\"{u}r Festk\"{o}rper- und Materialphysik, Technische Universit\"{a}t Dresden, 01069 Dresden, Germany}

\author{Riccardo Vocaturo}
\affiliation{Leibniz Institute for Solid State and Materials Research, IFW Dresden, 01069 Dresden, Germany}

\author{Hao Zhang}
\affiliation{School of Physics and Information Technology, Shaanxi Normal University, Xi'an 710119, China}

\author{\textcolor{red}{Qingxin Dong}}
\affiliation{\textcolor{red}{Beijing National Laboratory for Condensed Matter Physics and Institute of Physics, Chinese Academy of Sciences, Beijing 100190, China}}
\affiliation{\textcolor{red}{School of Physical Sciences, University of Chinese Academy of Sciences, Beijing 100190, China}}

\author{Man Li}
\thanks{lmrucphys@ruc.edu.cn}
\affiliation{School of Information Network Security, People's Public Security University of China, Beijing 100038, China}

\author{\textcolor{black}{Pengfei Ding}}
\affiliation{\textcolor{black}{School of Physics, Key Laboratory of Quantum State Construction and Manipulation (Ministry of Education), and Beijing Key Laboratory of Opto-Electronic Functional Materials $\&$ Micronano Devices, Renmin University of China, Beijing 100872, China}}

\author{\textcolor{black}{Erjian Cheng}}
\thanks{\textcolor{red}{Erjian.Cheng@cpfs.mpg.de}}
\affiliation{\textcolor{black}{Max Planck Institute for Chemical Physics of Solids, 01187 Dresden, Germany}}

\author{\textcolor{red}{Zhiguang Liao}}
\affiliation{\textcolor{black}{School of Physics, Key Laboratory of Quantum State Construction and Manipulation (Ministry of Education), and Beijing Key Laboratory of Opto-Electronic Functional Materials $\&$ Micronano Devices, Renmin University of China, Beijing 100872, China}}

\author{\textcolor{red}{Yu Zhang}}
\affiliation{\textcolor{black}{School of Physics, Key Laboratory of Quantum State Construction and Manipulation (Ministry of Education), and Beijing Key Laboratory of Opto-Electronic Functional Materials $\&$ Micronano Devices, Renmin University of China, Beijing 100872, China}}

\author{\textcolor{red}{Junfa Lin}}
\affiliation{\textcolor{black}{School of Physics, Key Laboratory of Quantum State Construction and Manipulation (Ministry of Education), and Beijing Key Laboratory of Opto-Electronic Functional Materials $\&$ Micronano Devices, Renmin University of China, Beijing 100872, China}}

\author{Reza Firouzmandi}
\affiliation{Leibniz Institute for Solid State and Materials Research, IFW Dresden, 01069 Dresden, Germany}

\author{Vilmos Kocsis}
\affiliation{Leibniz Institute for Solid State and Materials Research, IFW Dresden, 01069 Dresden, Germany}

\author{Laura T. Corredor}
\affiliation{Leibniz Institute for Solid State and Materials Research, IFW Dresden, 01069 Dresden, Germany}

\author{\textcolor{red}{Yurii Prots}}
\affiliation{\textcolor{black}{Max Planck Institute for Chemical Physics of Solids, 01187 Dresden, Germany}}

\author{\textcolor{red}{Oleksandr Suvorov}}
\affiliation{Leibniz Institute for Solid State and Materials Research, IFW Dresden, 01069 Dresden, Germany}

\author{\textcolor{red}{Anupam Jana}}
\affiliation{\textcolor{red}{CNR-IOM Istituto Officina dei Materiali, Area Science Park, 34149 Trieste, Italy}}
\affiliation{\textcolor{red}{International Centre for Theoretical Physics (ICTP), Str. Costiera 11, 34151 Trieste, Italy}}

\author{\textcolor{red}{Jun Fujii}}
\affiliation{\textcolor{red}{CNR-IOM Istituto Officina dei Materiali, Area Science Park, 34149 Trieste, Italy}}

\author{\textcolor{red}{Ivana Vobornik}}
\affiliation{\textcolor{red}{CNR-IOM Istituto Officina dei Materiali, Area Science Park, 34149 Trieste, Italy}}

\author{Oleg Janson}
\affiliation{Leibniz Institute for Solid State and Materials Research, IFW Dresden, 01069 Dresden, Germany}

\author{Wenliang Zhu}
\thanks{wlzhu@snnu.edu.cn}
\affiliation{School of Physics and Information Technology, Shaanxi Normal University, Xi'an 710119, China}
\affiliation{\textcolor{red}{State Key Laboratory of Precision Blasting, Jianghan University, Wuhan 430056, China}}

\author{Jeroen van den Brink}
\affiliation{Leibniz Institute for Solid State and Materials Research, IFW Dresden, 01069 Dresden, Germany}
\affiliation{Institute for Theoretical Physics and W\"{u}rzburg-Dresden Cluster of Excellence ct.qmat, Technische Universit\"{a}t Dresden, 01069 Dresden, Germany}

\author{Cornelius Krellner}
\affiliation{Kristall- und Materiallabor, Physikalisches Institut, Goethe-Universit\"{a}t Frankfurt, Max-von-Laue Strasse 1, 60438 Frankfurt am Main, Germany}

\author{Minghu Pan}
\affiliation{School of Physics and Information Technology, Shaanxi Normal University, Xi'an 710119, China}
\affiliation{School of Physics, Huazhong University of Science and Technology, Wuhan 430074, China}

\author{\textcolor{red}{Bosen Wang}}
\affiliation{\textcolor{red}{Beijing National Laboratory for Condensed Matter Physics and Institute of Physics, Chinese Academy of Sciences, Beijing 100190, China}}
\affiliation{\textcolor{red}{School of Physical Sciences, University of Chinese Academy of Sciences, Beijing 100190, China}}

\author{\textcolor{red}{Tianlong Xia}}
\affiliation{\textcolor{black}{School of Physics, Key Laboratory of Quantum State Construction and Manipulation (Ministry of Education), and Beijing Key Laboratory of Opto-Electronic Functional Materials $\&$ Micronano Devices, Renmin University of China, Beijing 100872, China}}

\author{\textcolor{red}{Jinguang Cheng}}
\affiliation{\textcolor{red}{Beijing National Laboratory for Condensed Matter Physics and Institute of Physics, Chinese Academy of Sciences, Beijing 100190, China}}
\affiliation{\textcolor{red}{School of Physical Sciences, University of Chinese Academy of Sciences, Beijing 100190, China}}

\author{\textcolor{black}{Shancai Wang}}
\affiliation{\textcolor{black}{School of Physics, Key Laboratory of Quantum State Construction and Manipulation (Ministry of Education), and Beijing Key Laboratory of Opto-Electronic Functional Materials $\&$ Micronano Devices, Renmin University of China, Beijing 100872, China}}

\author{\textcolor{black}{Claudia Felser}}
\affiliation{\textcolor{black}{Max Planck Institute for Chemical Physics of Solids, 01187 Dresden, Germany}}

\author{Bernd B\"{u}chner}
\affiliation{Leibniz Institute for Solid State and Materials Research, IFW Dresden, 01069 Dresden, Germany}
\affiliation{Institut f\"{u}r Festk\"{o}rper- und Materialphysik, Technische Universit\"{a}t Dresden, 01069 Dresden, Germany}

\author{Sergey Borisenko}
\affiliation{Leibniz Institute for Solid State and Materials Research, IFW Dresden, 01069 Dresden, Germany}

\author{\textcolor{red}{Rong Yu}}
\affiliation{\textcolor{black}{School of Physics, Key Laboratory of Quantum State Construction and Manipulation (Ministry of Education), and Beijing Key Laboratory of Opto-Electronic Functional Materials $\&$ Micronano Devices, Renmin University of China, Beijing 100872, China}}

\author{Denis V. Vyalikh}
\thanks{denis.vyalikh@dipc.org}
\affiliation{Donostia International Physics Center (DIPC), 20018 Donostia-San Sebasti\'{a}n, Spain}
\affiliation{IKERBASQUE, Basque Foundation for Science, 48011 Bilbao, Spain}

\author{Alexander Fedorov}
\thanks{a.fedorov@ifw-dresden.de}
\affiliation{Leibniz Institute for Solid State and Materials Research, IFW Dresden, 01069 Dresden, Germany}
\affiliation{Helmholtz-Zentrum Berlin f{\"u}r Materialien und Energie, Albert-Einstein-Stra{\ss}e 15, 12489 Berlin, Germany}
\affiliation{Joint Laboratory ``Functional Quantum Materials" at BESSY II, 12489 Berlin, Germany}

\begin{abstract}
  Applying angle-resolved photoemission spectroscopy and density functional theory calculations, we present compelling spectroscopic evidence demonstrating the intertwining and mutual interaction between the Kondo and kagome sublattices in heavy-fermion intermetallic compound YbV$_6$Sn$_6$. We reveal the Yb 4$f$-derived states near the Fermi level, along with the presence of bulk kagome bands and topological surface states. We unveil strong interactions between the 4$f$ and itinerant electrons, where the kagome bands hosting the Dirac fermions and van Hove singularities predominate. Such findings are well described using a $c$-$f$ hybridization model.
  \textcolor{black}{On the other hand, \textcolor{red}{our systematic characterization of magnetic properties} demonstrates an unusually enhanced antiferromagnetic ordering, \textcolor{red}{where the kagome-derived van Hove singularities near $E_F$ play a vital role in determining the unconventional nature of the Ruderman-Kittel-Kasuya-Yosida interaction and Kondo coupling. These unique kagome-state-mediated exchange interactions have never been reported before and could lead to} a novel phase diagram and various quantum critical behaviors in YbV$_6$Sn$_6$ and its siblings.}
  Our results not only expand the family of exotic quantum phases entangled with kagome structure to the strongly correlated regime, \textcolor{black}{but also establish YbV$_6$Sn$_6$ as an unprecedented platform to explore \textcolor{red}{unconventional many-body physics} beyond the standard Kondo picture.}
\end{abstract}

\maketitle

Exploring the mutual interactions among correlated states, magnetism, and topological phenomena is at the forefront of condensed-matter research \cite{DengY2020Science,SerlinM2020,ZhangP2018Science,ZhuS2020,GuQQ2023PDW,WangJ2023PDW}.
Kagome lattices are emerging platforms for studying this entanglement. A variety of electronic instabilities and nontrivial band topologies, such as Chern-gapped Dirac fermion \cite{YinJX2020Tb,JiangS2021Tb}, charge density wave \cite{Ortiz2019PRM,Ortiz2020PRL,YinQ2021,Ortiz2021PRM}, nematicity \cite{YangH2022CsTi3Bi5,LiH2023CsTi3Bi5}, superconductivity \cite{Ortiz2020PRL,YinQ2021,Ortiz2021PRM,YangH2022CsTi3Bi5}, and pair density wave \cite{ChenH2021PDW} have been observed.
Due to the unique lattice geometry, three typical band features are identified, including the flat band (FB), Dirac point (DP), and van Hove singularities (vHSs) \cite{LiM2021YMn6Sn6}.
\textcolor{black}{The essential role of kagome band structure in the formation of these symmetry-breaking states has been extensively studied \cite{YinJX2019CSS,ZengCG2022CoSn,TanHX2021,DennerMM2021,WenHH2021,Lou2022CVS,TengX2023FeGeNP,HeJF2023}.}


The discovery of heavy-fermion compound YbV$_6$Sn$_6$ offers a precious opportunity to study kagome physics in the presence of strongly correlated 4$f$ states \cite{JiaS2023Yb}. In the framework of the Kondo effect, when the $f$ electrons are arranged in a Kondo lattice, the strong interactions between itinerant conduction electrons and localized $f$ electrons lead to band reconstructions and opening of hybridization gaps, namely, the $c$-$f$ hybridization \cite{ColemanP2007,Vyalikh2010PRL,Vyalikh2011PRL}.
This gives rise to strongly renormalized quasiparticles with huge effective masses, called heavy fermions \cite{Stewart1984,Stewart2001}.
\textcolor{black}{Despite the tremendous efforts devoted to other intermetallic $R$V$_6$Sn$_6$ ($R$ = Y, Gd--Tm, Lu) \cite{Pokharel2021,LeeJ2022,Rosenberg2022,ChengJG2022,Pokharel2022}, the interplay of the correlation effects from the rare-earth sublattice and those associated with the exotic electronic states of V-based kagome network remains unexplored.}
With the adjacent triangular Kondo lattice and kagome lattice layers, \textcolor{black}{YbV$_6$Sn$_6$ provides a unique platform} to study the interplay between strong 4$f$ electronic correlations and kagome band structures. It is therefore desirable to explore spectroscopic evidence for these two sublattices and their potential coupling.

\textcolor{black}{In parallel, an antiferromagnetic ordering was observed in YbV$_6$Sn$_6$ below $\sim$0.4 K, and its suppression by a weak magnetic field induces the non-Fermi-liquid behavior \cite{JiaS2023Yb}, signifying the proximity of YbV$_6$Sn$_6$ to a potential quantum critical point (QCP). Further delving into the origin of magnetic ground state would shed light on the quantum criticality in a kagome metal, which still remains largely unknown.}

\textcolor{black}{By employing angle-resolved photoemission spectroscopy (ARPES) and density functional theory (DFT) calculations, here we uncover the entanglement of Kondo and kagome sublattices in YbV$_6$Sn$_6$--the strong hybridizations between Yb 4$f$ states close to $E_F$ and kagome bands constituting the DPs and vHSs.
Intriguingly, an unexpected enhancement of the antiferromagnetic order is observed. \textcolor{red}{We suggest that the unconventional nature of the underlying Ruderman-Kittel-Kasuya-Yosida (RKKY) and Kondo interactions originates from their being mediated primarily by the kagome electrons, where the enhanced density of states (DOS) and electron correlations associated with vHSs play a crucial role.}} 

\begin{figure}[b]
  \setlength{\abovecaptionskip}{-0.03cm}
  \setlength{\belowcaptionskip}{-0.2cm}
  \begin{center}
  \includegraphics[trim = 0mm 0mm 0mm 0mm, clip=true, width=1.0\columnwidth]{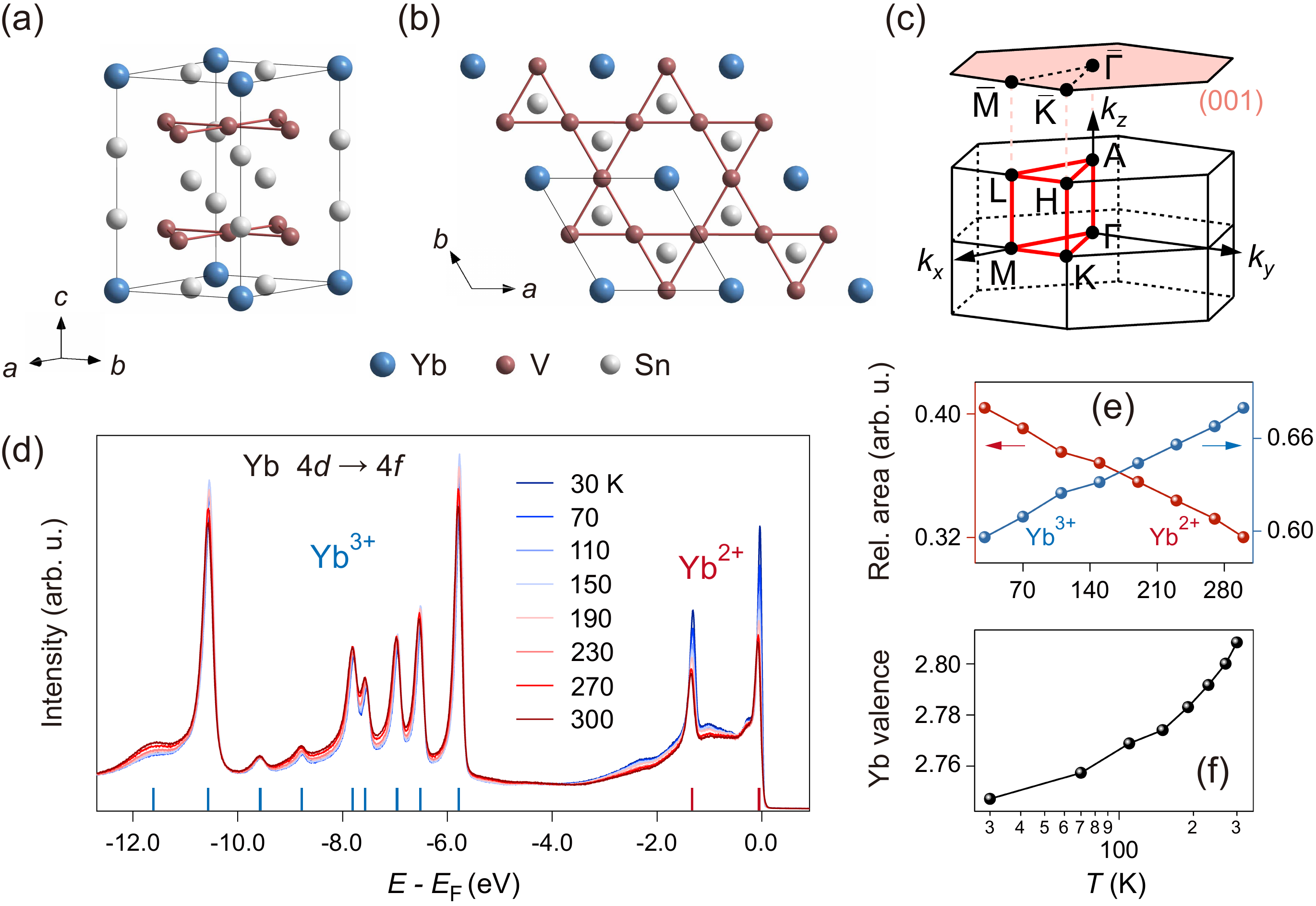}
  \end{center}
  \caption{
  (a),(b) Side and top views of the YbV$_6$Sn$_6$ crystal structure, respectively.
  (c) Bulk and (001)-projected surface BZs.
  (d) Temperature-dependent angle-integrated photoemission intensity ($h\nu$ = 182 eV). 
  The Yb$^{2+}$ and Yb$^{3+}$ components are marked out by the red and blue bars, respectively.
  \textcolor{black}{(e) Temperature-dependent relative peak areas of Yb$^{2+}$ and Yb$^{3+}$ features. Before the normalization, a Shirley-type background has been subtracted from both components.
  (f) Temperature dependence of the estimated Yb valence.}
  }
\end{figure}

High-quality YbV$_6$Sn$_6$ single crystals were grown by the self-flux method \cite{JiaS2023Yb,SM}. YbV$_6$Sn$_6$ crystallizes in a $P$6/$mmm$ HfFe$_6$Ge$_6$-type structure \cite{Fredrickson2008}, \textcolor{red}{featuring the Yb-based triangular and V-based kagome sublattices [Figs. 1(a) and 1(b)].}
The bulk and (001)-projected surface Brillouin zones (BZs) are depicted in Fig. 1(c). The occurrence of intermediate valence in rare-earth intermetallic materials has been suggested as characteristic of the Kondo effect \cite{Varma1976valence}.
\textcolor{black}{To examine the 4$f$ occupation in YbV$_6$Sn$_6$, we measure the temperature-dependent angle-integrated photoemission spectra on the Sn termination [Fig. 1(d)].} Under the Yb 4$d$$\rightarrow$4$f$ resonant photon energy \cite{Johansson1980}, we reveal the Yb$^{2+}$ (4$f^{14}$$\rightarrow$4$f^{13}$ transition) and Yb$^{3+}$ (4$f^{13}$$\rightarrow$4$f^{12}$ transition) components, manifesting as doublet (4$f_{7/2}$ and 4$f_{5/2}$) and multiplet, respectively.
The extracted \textcolor{black}{relative peak areas} of Yb$^{2+}$ and Yb$^{3+}$ features clearly show the temperature evolution of 4$f$ occupation, as presented in Fig. 1(e).
\textcolor{black}{Accordingly, the Yb valence can be estimated [Fig. 1(f), \textcolor{red}{see estimation process and more discussion} in Sec. 2 of Supplemental Material (SM)].
Their temperature dependences are similar to other Yb-based heavy-fermion and valence-fluctuating systems, like YbRh$_2$Si$_2$ \cite{Vyalikh2015PRX}, YbInCu$_4$ \cite{Okusawa1996YbInCu4,SatoH2004YbInCu4}, and YbAl$_3$ \cite{ShenKM2017YbAl3}.}
The Kondo lattice behavior of YbV$_6$Sn$_6$ is also embodied in electrical resistivity measurements (Fig. S1), where the zero-field $\rho$($T$) resembles that of many other heavy-fermion compounds \cite{Canfield1999,Canfield2004,Canfield2013}.

\begin{figure}[b]
  \setlength{\abovecaptionskip}{-0.03cm}
  \setlength{\belowcaptionskip}{-0.2cm}
  \begin{center}
  \includegraphics[trim = 0mm 0mm 0mm 0mm, clip=true, width=1.0\columnwidth]{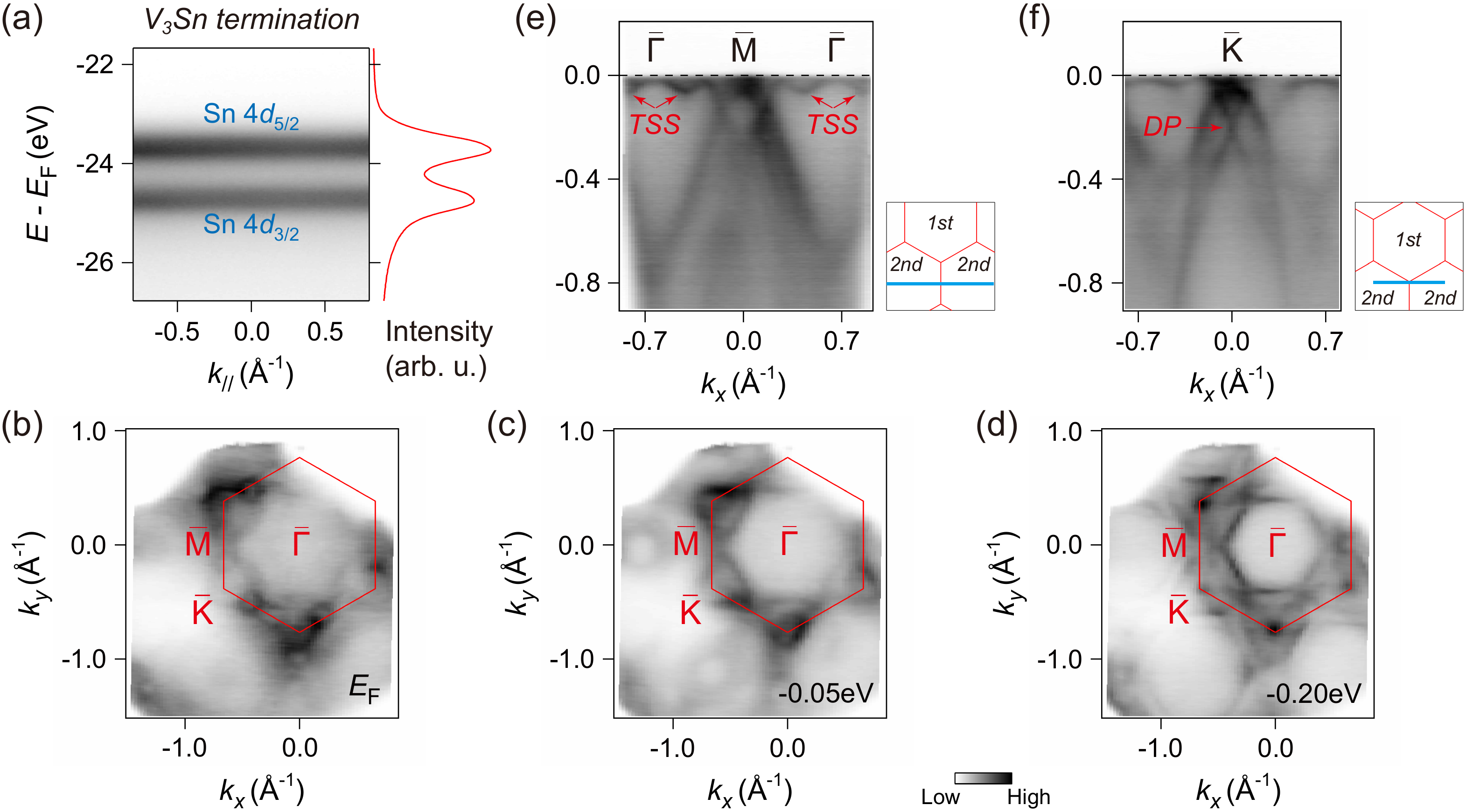}
  \end{center}
  \caption{
  (a) X-ray photoelectron spectroscopy spectrum ($h\nu$ = 100 eV) (left) and corresponding integrated energy distribution curve (right) measured on the V$_3$Sn termination.
  (b)-(d) Constant-energy ARPES images [$h\nu$ = 80 eV, linear horizontal (LH) polarization] from the V$_3$Sn termination at the energies of 0, -0.05, and -0.20 eV, respectively.
  (e),(f) Corresponding ARPES intensity plots \textcolor{black}{($T$ $\approx$ 2 K)} around $\bar{M}$ and $\bar{K}$, respectively. The momentum locations are illustrated in the insets.
  The 50-eV photons were used to better reveal the TSSs and DP.
  }
\end{figure}

\begin{figure*}[t]
  \setlength{\abovecaptionskip}{-0.03cm}
  \setlength{\belowcaptionskip}{-0.2cm}
  \begin{center}
  \includegraphics[trim = 0mm 0mm 0mm 0mm, clip=true, width=1.8\columnwidth]{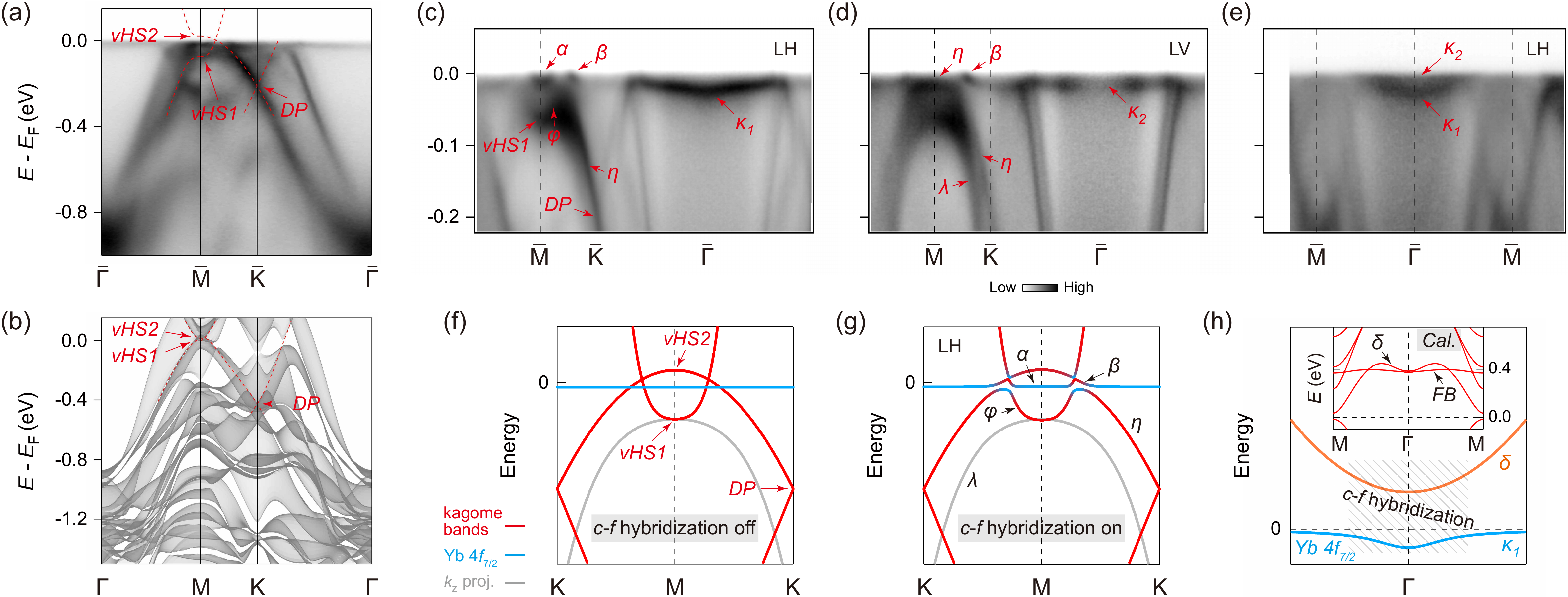}
  \end{center}
  \caption{
  (a) Experimental band structures of YbV$_6$Sn$_6$ along the $\bar{\Gamma}$-$\bar{M}$-$\bar{K}$-$\bar{\Gamma}$ lines \textcolor{black}{at 2 K.} The typical kagome bands are guided by the red dashed curves (without including the effect of $c$-$f$ hybridization).
  (b) DFT calculated bulk bands with the integration over entire $k_z$. The contribution of the Yb 4$f$ shell is not included.
  \textcolor{black}{(c),(d)} ARPES intensity plots along the $\bar{\Gamma}$-$\bar{K}$-$\bar{M}$ directions \textcolor{black}{at 2 K} recorded by 53-eV photons with LH and LV polarizations, \textcolor{black}{respectively.}
  \textcolor{black}{(e)} Same as \textcolor{black}{(c)} along the $\bar{\Gamma}$-$\bar{M}$ direction.
  \textcolor{black}{(f),(g)} Schematic band diagrams along the $\bar{K}$-$\bar{M}$-$\bar{K}$ directions without and with considering the $c$-$f$ hybridization, \textcolor{black}{respectively.} A simple band hybridization model is constructed.
  \textcolor{black}{(h)} Sketches of the hybridization between $\delta$ band and 4$f_{7/2}^{13}$ state at $\bar{\Gamma}$. Inset: Calculated bulk bands along the $M$-$\Gamma$-$M$ direction.
  }
\end{figure*}

To characterize the typical electronic structures of kagome [Fig. S2(a)] and Kondo [Fig. S2(b)] sublattices as well as their underlying interactions, we conduct ARPES experiments on YbV$_6$Sn$_6$. Based on the line shape of Sn 4$d$ core levels \cite{Kang2020FeSn} and no evident Yb 4$f$ surface signals \cite{Molodtsov2007fsurface} in our measurements, we obtained two surface terminations upon cleaving (V$_3$Sn and Sn). The corresponding band structures are shown in Figs. 2 and S3. Similar to GdV$_6$Sn$_6$ \cite{HeJF2021Gd} and TbV$_6$Sn$_6$ \cite{Sante2023TbNP}, the Sn termination shows Sn 4$d$ core levels from both bulk and surface Sn atoms [Fig. S3(a)], while the V$_3$Sn termination only shows the signals from bulk Sn atoms [Fig. 2(a)].
Although the ARPES mappings from Sn termination exhibit richer topologies [Figs. S3(b)-S3(d)] compared to V$_3$Sn termination [Figs. 2(b)-2(d)], the typical features of a kagome lattice, the corner-sharing triangular pockets centered at $\bar{K}$ \cite{Lou2022CVS}, are recognized on both terminations.


Figures 2(e) and S3(e) show the ARPES spectra along the $\bar{\Gamma}$-$\bar{M}$-$\bar{\Gamma}$ directions,
\textcolor{black}{two shallow electron bands are observed in the second BZ for both terminations,
where the intensity on the Sn termination is intrinsically weak (more pronounced) at higher (lower) photon energies due to the $k_z$ selection rules (Fig. S4).}
These bands resemble the topological surface states (TSSs) in GdV$_6$Sn$_6$ \cite{Hu2022Gd} (see Fig. S5 and Sec. 3 of SM for the slab calculations and more discussion).
We further measure the band dispersions near $\bar{K}$. As shown in Figs. 2(f) and S3(f), the linear bands cross each other at about -0.2 eV to form the kagome DP at $\bar{K}$. Such bulk kagome bands are native to the kagome lattice irrespective of surface termination.
Below we focus on the intrinsic bulk kagome band structures.



Figure 3(a) summarizes the experimental band dispersions along the $\bar{\Gamma}$-$\bar{M}$-$\bar{K}$-$\bar{\Gamma}$ lines. By comparing with the DFT calculations [Figs. 3(b) and S6], we obtain a good overall agreement and experimentally identify the typical band structure of the kagome sublattice.
Specifically, the vHS1 is observed at about -0.07 eV at $\bar{M}$; the DP at about -0.2 eV is revealed around $\bar{K}$; the upper branch of the DP disperses up to constitute the vHS2 slightly above $E_F$ at $\bar{M}$. Similar kagome bands near $E_F$ have also been identified in GdV$_6$Sn$_6$ \cite{Hu2022Gd} and TbV$_6$Sn$_6$ \cite{Rosenberg2022}.

We now proceed to explore the ARPES signatures of the Kondo sublattice and its potential intertwining with the kagome sublattice.
As shown in \textcolor{black}{Fig. 3(c)} (LH polarization), the bulk kagome bands identified before exhibit appreciable discontinuity when approaching the Yb 4$f_{7/2}^{13}$ states close to $E_F$, in sharp contrast to other $R$V$_6$Sn$_6$ compounds without 4$f$ levels near $E_F$, like GdV$_6$Sn$_6$ \cite{Hu2022Gd,HeJF2021Gd}, TbV$_6$Sn$_6$ \cite{Rosenberg2022}, and HoV$_6$Sn$_6$ \cite{HeJF2021Gd}. The electron-like vHS1 band ($\varphi$) and the Dirac band ($\eta$)
bend back to merge with each other below $E_F$ [see \textcolor{black}{Figs. S7(b)(i),(c)(i) and S8(b)(i),(c)(i)} under other photon energies]. A depletion of intensity is observed between the $\varphi$/$\eta$ band top and the two shallow bands ($\alpha$ and $\beta$), pointing to the opening of energy gaps. These facts suggest the presence of hybridizations between kagome bands and 4$f$ states, which can be well accounted for by the $c$-$f$ hybridization model \cite{Vyalikh2010PRL,Vyalikh2011PRL}.
As illustrated in \textcolor{black}{Fig. 3(f),} we sketch the original features along the $\bar{K}$-$\bar{M}$-$\bar{K}$ line \textcolor{black}{before the hybridization.} It is seen that both the vHS1 band and Dirac band (or vHS2 band) will interact with the dispersionless 4$f_{7/2}^{13}$ state. When the $c$-$f$ hybridization is included \textcolor{black}{[Fig. 3(g)]}, hybridization gaps open at the intersections. Specifically, the 4$f$-derived states disperse upward and merge with the upper unhybridized parts of kagome bands; the 4$f$ level merged with the vHS1 band further interacts with the vHS2 band slightly above $E_F$, leading to the $\alpha$ and $\beta$ bands; the lower unhybridized kagome bands bend back to form the $\varphi$/$\eta$ band top with strong 4$f$ admixtures [also see band structure cartoon in \textcolor{black}{Fig. S9(b)}].
\textcolor{red}{The validity of this model is further corroborated by our Cs-dosing experiments (Fig. S10).}
\textcolor{black}{Systematic temperature-dependent measurements \textcolor{red}{(Fig. S11)} demonstrate that such hybridizations occur already at high temperatures (see detailed discussion in \textcolor{red}{Sec. 5} of SM).}

We then record the ARPES spectra with linear vertical (LV) polarization. In \textcolor{black}{Fig. 3(d),} one sees that the $\beta$ band remains while the $\alpha$ and $\varphi$ bands are not present, in addition, the $\eta$ band disperses towards $E_F$ till being flattened around $\bar{M}$ [see \textcolor{black}{Figs. S7(b)(ii),(c)(ii) and S8(b)(ii),(c)(ii)} under other photon energies]. Accordingly, a gap opens between the $\beta$ and $\eta$ bands. \textcolor{black}{With the three-dimensionality of the electronic structure revealed by photon-energy-dependent measurements \textcolor{red}{(Fig. S12),} the vHS1 band (along $\bar{K}$-$\bar{M}$-$\bar{K}$ direction) would lie completely above the 4$f_{7/2}^{13}$ state in certain $k_z$ planes, as seen in Figs. 3(b) and S6. The projection from such $k_z$'s coexists with the one from $k_z$ = 0 due to the $k_z$-broadening effect \cite{Strocov2003kz}. Whereas, these two sets of $k_z$-projected vHS1 bands most likely have different orbital characters \cite{BHYan2023Sc}, making themselves separately visible under different photon polarizations (see Table S1 and \textcolor{red}{Sec. 7} of SM for the matrix element analysis).
In the case of LV polarization, as illustrated in Fig. S9(c), only the vHS2 band will interact with the 4$f$ state, resulting in the observations in Fig. 3(d).}

Next, we turn to the band structure around $\bar{\Gamma}$. In \textcolor{black}{Fig. 3(c),} the 4$f_{7/2}^{13}$ state ($\kappa_{\rm 1}$) is pushed slightly downwards at $\bar{\Gamma}$ \textcolor{red}{(see Fig. S12 for its bulk origin).} Similar phenomenon has been observed before in YbRh$_2$Si$_2$ and suggested to originate from the interactions between the 4$f$ states and an unoccupied electron band \cite{Vyalikh2010PRL}. Such hybridization can be described by the periodic Anderson model calculations \cite{Molodtsov2007fsurface} and has been evidenced by the extended quasiparticle lifetimes of the involved unoccupied states in time-resolved ARPES measurements of YbRh$_2$Si$_2$ \cite{Vyalikh2012trARPES}. In our case,
as suggested by the calculations [inset of \textcolor{black}{Fig. 3(h) and S6}], the corresponding unoccupied band could be the parabolic band ($\delta$) whose bottom degenerates with the kagome FB at $\sim$0.37 eV. \textcolor{black}{Figure 3(h)} depicts the hybridization between $\delta$ band and 4$f_{7/2}^{13}$ state, which is another indicator of the intertwined Kondo and kagome sublattices, because the $\delta$ extends to form the kagome band features at $M$ and $K$ \textcolor{black}{(Fig. S6).} Upon switching to LV polarization \textcolor{black}{[Fig. 3(d)],} instead of $\kappa_{\rm 1}$, a flat 4$f$ state ($\kappa_{\rm 2}$) without hybridization is revealed at $\bar{\Gamma}$ [also see \textcolor{black}{Fig. S8(a)(ii)}]. These results taken together indicate the presence of two split 4$f_{7/2}^{13}$ states around $\bar{\Gamma}$. This is further underpinned by the spectra along the $\bar{\Gamma}$-$\bar{M}$ direction [Fig. 3(e)], where the two split states are simultaneously detected (see \textcolor{black}{Figs. S7 and S8} for more data). Such observations provide direct evidence that the 4$f_{7/2}^{13}$ state undergoes the crystalline electric field splitting \cite{CEF1985} whose energy scale at BZ center is of $\sim$20 meV, which is comparable to the typical order of magnitude for crystalline electric field effects in 4$f$-based systems \cite{Vyalikh2010PRL}.

\begin{figure}[t]
  \setlength{\abovecaptionskip}{-0.03cm}
  \setlength{\belowcaptionskip}{-0.2cm}
  \begin{center}
  \includegraphics[trim = 0mm 0mm 0mm 0mm, clip=true, width=1.0\columnwidth]{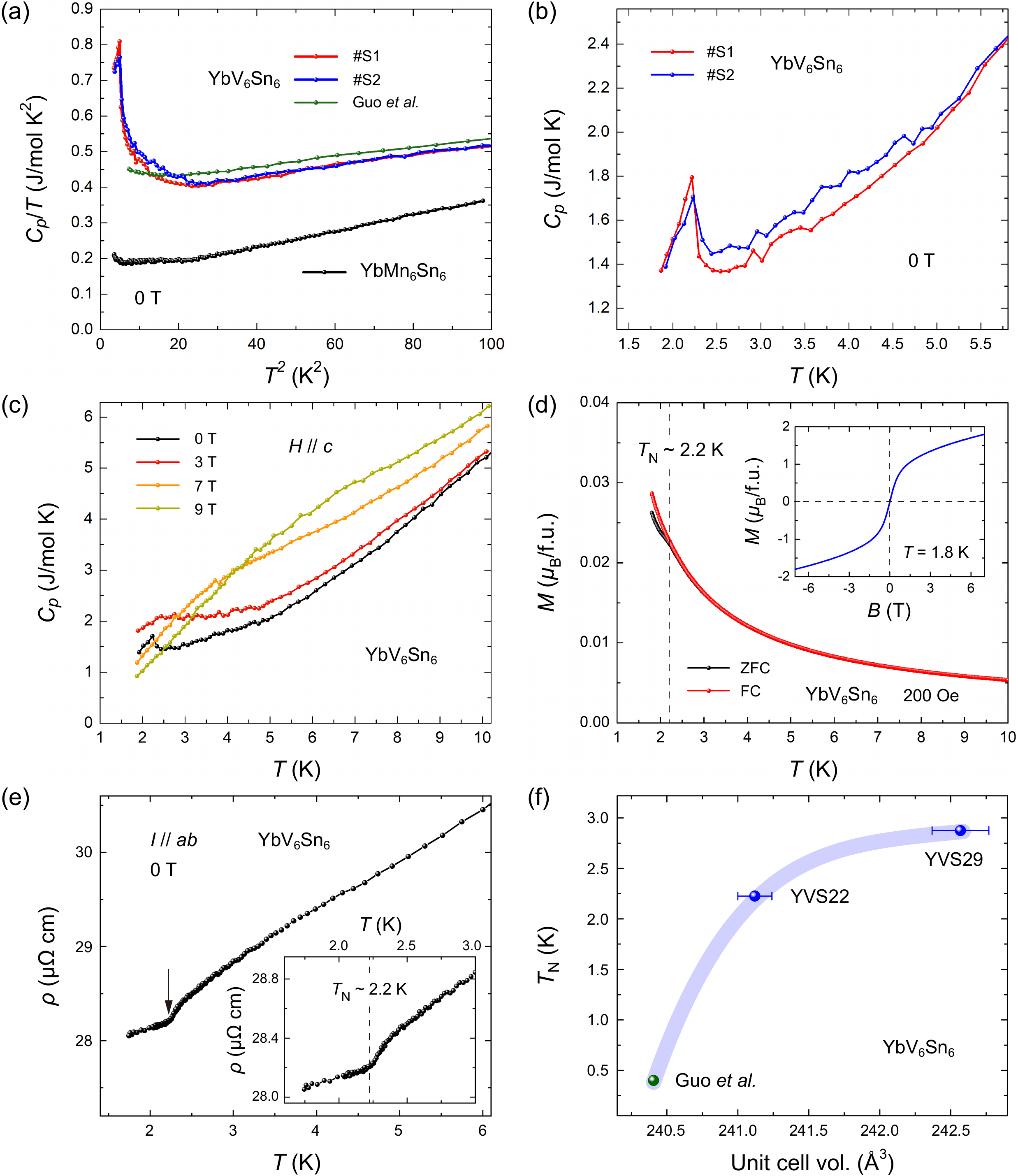}
  \end{center}
  \caption{\textcolor{black}{
  (a) Zero-field specific heat of YbV$_6$Sn$_6$ and YbMn$_6$Sn$_6$ plotted as $C_p$/$T$ vs $T^2$. The green curve is adopted from Ref. [\onlinecite{JiaS2023Yb}].
  (b) Low-temperature specific heat of YbV$_6$Sn$_6$ in zero field.
  (c) Temperature-dependent specific heat of YbV$_6$Sn$_6$ in different magnetic fields ($H$ $\parallel$ $c$).
  (d) Temperature dependence of zero-field-cooling and field-cooling magnetization of YbV$_6$Sn$_6$ ($H$ $\parallel$ $ab$). Inset: Field-dependent magnetization at 1.8 K ($H$ $\parallel$ $ab$).}
  \textcolor{red}{
  (e) Temperature-dependent zero-field resistivity of YbV$_6$Sn$_6$ ($I$ $\parallel$ $ab$). Inset shows temperature range close to $T_{\rm N}$.
  (f) $T_{\rm N}$ of YbV$_6$Sn$_6$ as a function of unit cell volume. The green dot is adopted from Ref. [\onlinecite{JiaS2023Yb}].}
  }
\end{figure}

\textcolor{black}{A recent study proposed that YbV$_6$Sn$_6$ is near a potential antiferromagnetic QCP \cite{JiaS2023Yb}. To gain insights into the magnetic ground state, we perform low-temperature heat capacity, magnetization, \textcolor{red}{and electrical resistivity} measurements. As presented in Fig. 4(a), the zero-field $C_p$/$T$ of our samples follows similar Fermi-liquid behavior as in Ref. [\onlinecite{JiaS2023Yb}] from $\sim$4 to 10 K, showing a comparable Sommerfeld coefficient ($\gamma$ $\sim$ 383 mJ/mol K$^2$). Interestingly, a pronounced increase below $\sim$4 K followed by a peak formation at $\sim$2.2 K are observed. The $\lambda$-shaped peak at $\sim$2.2 K in $C_p$ [Fig. 4(b)] indicates a second-order phase transition. The evolution of this peak under the magnetic field [Fig. 4(c)] is similar to that of the 0.4-K peak in Ref. [\onlinecite{JiaS2023Yb}]. \textcolor{red}{Further, our magnetization [Fig. 4(d)] and electrical resistivity [Fig. 4(e) and Sec. 9 of SM] measurements} confirm that the nature of this transition is indeed an antiferromagnetically ordered state. \textcolor{red}{These observations demonstrate that the previously reported} antiferromagnetism has been notably enhanced in our sample \textcolor{red}{(labeled as YVS22).}}

\textcolor{black}{The proximity to QCP renders the antiferromagnetic order sensitive to the extrinsic parameters like pressure, magnetic field, or chemical substitution \cite{Schuberth2016}. \textcolor{red}{Given the observed lattice expansion in YVS22 (see Fig. S15 for the x-ray diffraction data) compared to Guo \emph{et al.}'s \cite{JiaS2023Yb}, it is inferred that the revealed enhancement could arise from the negative chemical pressure effect. To investigate whether $T_{\rm N}$ can be further increased, we have slightly modified the growth conditions. It turns out that slightly elevated growth temperatures result in an increase of $T_{\rm N}$ to $\sim$2.9 K [Fig. S16(a), labeled as YVS29]. Meanwhile, the lattice of YVS29 (Table S2) is further expanded compared to YVS22. Accordingly, the variation of $T_{\rm N}$ with the unit cell volume is summarized in Fig. 4(f). Although the origin of the lattice expansion remains unclear and warrants further investigation, the present results indicate an intrinsic link between antiferromagnetic order and lattice parameters. Furthermore, our high-pressure studies of YVS29 reveal that the antiferromagnetic order is progressively weakened with increasing hydrostatic pressure [Figs. S16(b)-S16(d)]. Taken together, these results demonstrate that the antiferromagnetism in YbV$_6$Sn$_6$ strengthens with lattice expansion and diminishes under lattice compression, establishing YbV$_6$Sn$_6$ as a compelling system for realizing a pressure-induced QCP. An alternative approach to accessing the QCP would be through fine-tuned Mn substitution on V sites,} because the Mn sublattices order ferromagnetically while the Yb atoms have no magnetic moment in YbMn$_6$Sn$_6$ \cite{Eichenberger2017,Magnette2018}.
The obtained $\gamma$ $\sim$ 144 mJ/mol K$^2$ [Fig. 4(a)] and intermediate valence behavior \textcolor{red}{(Fig. S17)} evidence the presence of heavy-fermion state in YbMn$_6$Sn$_6$ (see more discussion in \textcolor{red}{Sec. 12} of SM). Our findings establish a robust heavy-fermion kagome system irrespective of the rich magnetic ground states.}

\textcolor{red}{In a Kondo lattice system, the magnetic ground state is determined by the competition between RKKY and Kondo interactions, whose energy scales are characterized by $T_{\rm RKKY}$$\propto$$J^{2}g(E_F)$ and $T_{\rm K}$$\propto$exp(-1/$J$$g$($E_F$)), respectively, where $g$($E_F$) is the conduction-band DOS at $E_F$ and $J$$\propto$$V^{2}$/($E_F$-$E_f$) is the magnetic exchange coupling between conduction electrons and $f$ electrons, depending on the $c$-$f$ hybridization ($V$) and the energy location of $f$ level ($E_f$). According to the Doniach diagram \cite{Doniach1977}, YbV$_6$Sn$_6$ is situated in the intermediate $J$ regime. In a conventional Kondo picture, pressurizing such Yb-based compounds would suppress the valence fluctuations and readily drive the system towards the 4$f^{13}$ configuration, resulting in more localized 4$f$ states and weaker $J$ \cite{YbNiSn1995}. Consequently, $T_{\rm RKKY}$ and $T_{\rm K}$ decrease with increasing positive pressure, and $T_{\rm N}$ would first increase and then decrease. Accordingly, the negative pressure would enhance $J$, and the increasingly screened magnetic moments would lead to the decrease of $T_{\rm N}$. However, this conventional Kondo paradigm fails to account for our observations, indicating an exotic origin. In Fig. S19, our DFT calculations reveal that the vHS1 and vHS2 remain close to $E_F$ under various hydrostatic pressures, giving rise to a large DOS around $E_F$ and enhanced many-body interactions, which imply that much more conduction states are involved in the $c$-$f$ hybridization in YbV$_6$Sn$_6$ compared to other systems. These abundant kagome electron channels can promote the valence fluctuations and stabilize the intermediate valence states. Thereby, the Yb valence would evolve smoothly and slowly under pressure, rendering the pressure dependence of $J$ predominantly governed by $V$. In this context, $J$ would be enhanced (weakened) by the lattice compression (lattice expansion) due to the increased (reduced) overlap between the wavefunctions of conduction states and $f$ states. Thus, $T_{\rm RKKY}$ and $T_{\rm K}$ of YbV$_6$Sn$_6$ would increase (decrease) with increasing positive pressure (negative pressure), and the increasingly (less) screened magnetic moments result in the decrease (increase) of $T_{\rm N}$ (see more detailed discussion in Sec. 13 of SM). Such kagome-state-mediated RKKY and Kondo interactions, intertwined with the geometric frustration, underpin a landscape of unconventional many-body physics beyond the standard Kondo framework.}


\textcolor{black}{In summary, we have revealed a strong entanglement between the Kondo and kagome sublattices in YbV$_6$Sn$_6$. \textcolor{red}{The unique kagome-mediated RKKY and Kondo interactions lead to an emergent antiferromagnetic ground state.} Our results highlight YbV$_6$Sn$_6$ as a promising system that can exhibit novel quantum criticality and richer Kondo physics.}

\begin{acknowledgments}
\textcolor{red}{We would like to thank Rui-Zhen Huang and Anmin Zhang for inspiring discussions.}
This work was supported by the Deutsche Forschungsgemeinschaft (DFG) under Grant SFB 1143 (project C04), the W{\"u}rzburg-Dresden Cluster of Excellence on Complexity and Topology in Quantum Matter -- \emph{ct.qmat} (EXC 2147, project ID 390858490), \textcolor{red}{the DFG through the TRR 288 (project ID 422213477, project A03),} the National Key R\&D Program of China (Grants No. 2022YFA1403100, No. 2022YFA1403101, and \textcolor{red}{No. 2022YFA1403800}), the National Natural Science Foundation of China (Grants No. 12204297, No. 12204536, and \textcolor{red}{No. 12274455}), the Fundamental Research Funds for the Central Universities, and the Research Funds of People's Public Security University of China (PPSUC) (Grant No. 2023JKF02ZK09). \textcolor{red}{W. L. Z. also acknowledges the support from the State Key Laboratory of Precision Blasting, Jianghan University (Grant No. PBSKL25A07).} \textcolor{black}{E. J. C. acknowledges the financial support from the Alexander von Humboldt Foundation. L. T. C. is funded by the DFG (project-id 456950766).} B. B. and S. B. acknowledge the support from the BMBF via project UKRATOP.
\textcolor{red}{The high-pressure measurements were partially supported by the Cubic Anvil Cell station of Synergic Extreme Condition User Facility (SECUF).}

\end{acknowledgments}




\end{document}